\documentclass[letterpaper]{article} 
\usepackage{aaai25}
\usepackage{times}  
\usepackage{helvet}  
\usepackage{courier}  
\usepackage[hyphens]{url}  
\usepackage{graphicx} 
\usepackage{natbib}  
\usepackage{caption} 
\usepackage{algorithm}
\usepackage{algorithmic}
\usepackage{subcaption}
\usepackage{booktabs}
\usepackage{multirow}
\usepackage{graphicx}
\usepackage{subfig}
\usepackage{caption}
\usepackage{dcolumn}
\usepackage{amsfonts}
\usepackage{amsmath}
\usepackage{tcolorbox}
\usepackage{appendix}
\usepackage{newfloat}
\usepackage{listings}
\DeclareCaptionStyle{ruled}{labelfont=normalfont,labelsep=colon,strut=off} 
\floatstyle{ruled}
\newfloat{listing}{tb}{lst}{}
\floatname{listing}{Listing}
\newcolumntype{R}[1]{>{\centering\arraybackslash}p{#1}}

\usepackage[nolist]{acronym}
\usepackage[capitalize]{cleveref}
\begin{acronym}
  \acro{LLM}{large language model}
  \acro{MLLM}{multimodal large language model}
  \acro{ADA-Steal}[\textsc{ADA-Steal}]{Adversarial Domain Alignment}
\end{acronym}

\title{Medical Multimodal Model Stealing Attacks via Adversarial Domain Alignment}
\author{
    Yaling Shen\textsuperscript{\rm 1,2,3}\equalcontrib,
    Zhixiong Zhuang\textsuperscript{\rm 1,4}\equalcontrib\thanks{{Corresponding author.}},
    Kun Yuan\textsuperscript{\rm 2,3,5},
    Maria-Irina Nicolae\textsuperscript{\rm 1}, \\
    Nassir Navab\textsuperscript{\rm 2}, 
    Nicolas Padoy\textsuperscript{\rm 5,6}, 
    Mario Fritz\textsuperscript{\rm 7}
}
\affiliations{
    \textsuperscript{\rm 1}Bosch Center for Artificial Intelligence, Germany \\
    \textsuperscript{\rm 2}Technical University of Munich, Germany \\
    \textsuperscript{\rm 3}Munich Center for Machine Learning, Germany\\
    \textsuperscript{\rm 4}Saarland University, Germany\\
    \textsuperscript{\rm 5}University of Strasbourg, France \\
    \textsuperscript{\rm 6}IHU Strasbourg, France \\
    \textsuperscript{\rm 7}CISPA Helmholtz Center for Information Security, Germany \\
    \{zhixiong.zhuang, irina.nicolae\}@bosch.com, \{yaling.shen, kun.yuan, nassir.navab\}@tum.de, \\
    npadoy@unistra.fr, fritz@cispa.de
}

\usepackage{bibentry}

\begin{document}

\maketitle

\begin{abstract}
Medical \acp{MLLM} are becoming an instrumental part of healthcare systems, assisting medical personnel with decision making and results analysis.
Models for radiology report generation are able to interpret medical imagery, thus reducing the workload of radiologists.
As medical data is scarce and protected by privacy regulations, medical \acp{MLLM} represent valuable intellectual property. However, these assets are potentially vulnerable to model stealing, where attackers aim to replicate their functionality via black-box access.
So far, model stealing for the medical domain has focused on classification; however, existing attacks are not effective against \acp{MLLM}.
In this paper, we introduce \ac{ADA-Steal}, the first stealing attack against medical \acp{MLLM}.
\ac{ADA-Steal} relies on natural images, which are public and widely available, as opposed to their medical counterparts.
We show that data augmentation with adversarial noise is sufficient to overcome the data distribution gap between natural images and the domain-specific distribution of the victim \ac{MLLM}.
Experiments on the \textsc{IU X-Ray} and \textsc{MIMIC-CXR} radiology datasets demonstrate that \acl{ADA-Steal} enables attackers to steal the medical \ac{MLLM} without any access to medical data.
\end{abstract}

%

\section{Introduction}
In recent years, the development of medical \acfp{MLLM} has garnered widespread attention due to their potential to revolutionize healthcare. These models could support clinical decision-making~\cite{seenivasan2022surgical,chen2024vs,peskaVLP}, enhance diagnostic accuracy~\cite{yuan2024advancing,chexagent}, and promote equitable distribution of medical resources~\cite{jia2024medpodgpt,yuan2023learning,zhang2023biomedgpt}.
One of the most important use cases is radiology report generation, where the medical \ac{MLLM} takes a radiology image (e.g., chest X-ray) as input and generates a detailed diagnostic report.
Since medical data is usually not publicly available and medical expertise is scarce, a well-performing medical \acp{MLLM} becomes valuable intellectual property (IP).
However, these models are potentially vulnerable to model stealing attacks~\citep{steal-api,knockoff}, which replicate the functionality of a machine learning model through black-box access.
This threat is particularly significant in the medical field because a duplicated model could conflict with IP and privacy regulations, and could also facilitate further attacks.
By designing a payload on the copied model, malicious actors can then attack the original model, e.g., with a transfer jailbreak attack~\cite{huang2024crossmodality} to compel a medical \acp{MLLM} to output fraudulent or fabricated medical information.

\begin{figure}[t]
\centering
\includegraphics[width=1.0\columnwidth]{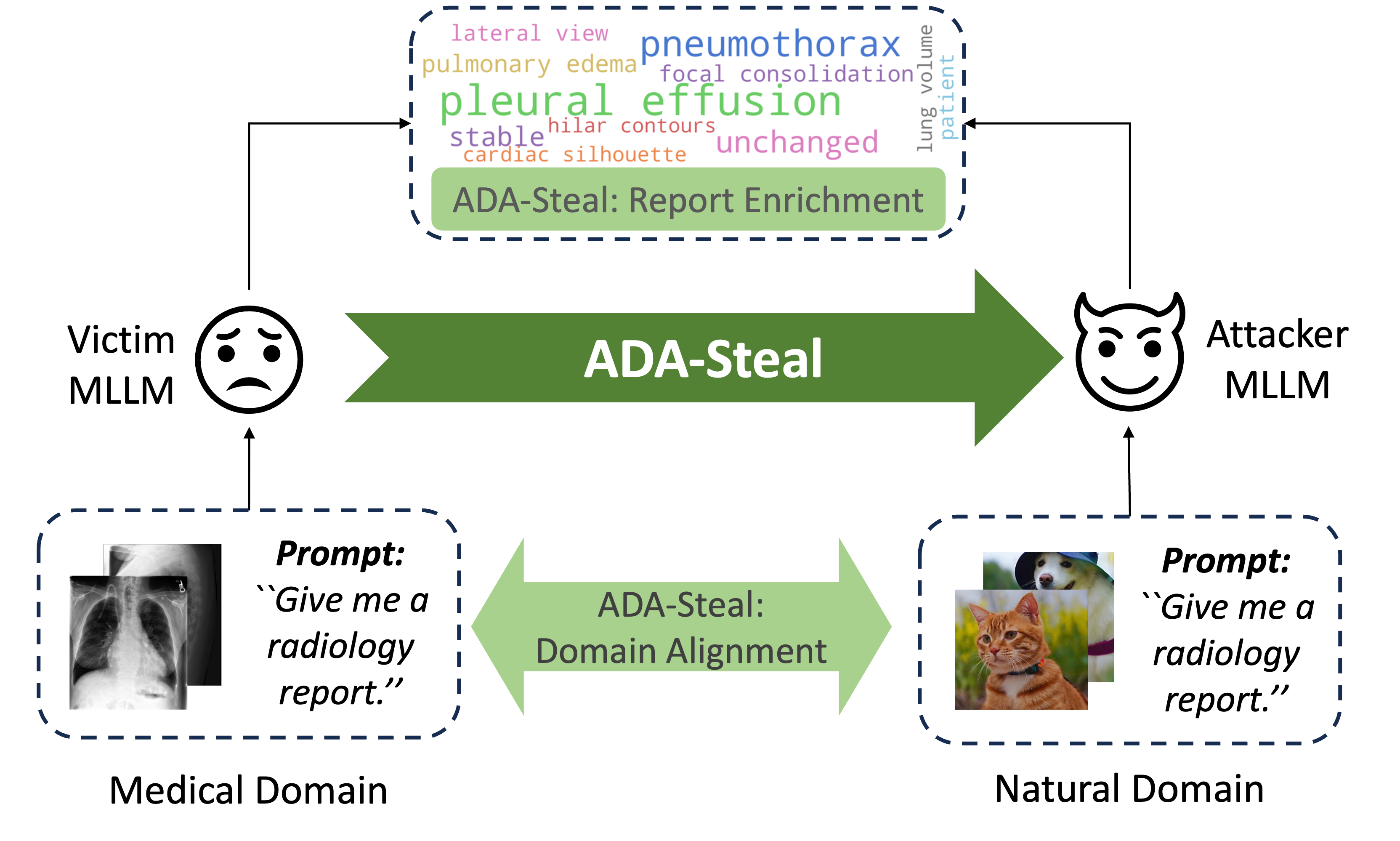} 
\caption{\ac{ADA-Steal} trains an attacker \ac{MLLM} to replicate a victim \ac{MLLM} for radiology report generation from natural images by first enriching the reports and then aligning the attacker distribution to the medical domain.}
\label{fig: teaser}
\end{figure}

Model theft typically involves two steps: (i) creating a transfer dataset by querying the victim model with public or synthetic data~\citep{knockoff, dfme} to obtain pseudo-labels; and (ii) training the attacker model using these pseudo-labels as ground truth.
Nevertheless, current model stealing methods present significant limitations when applied to medical \acp{MLLM}.
First, due to patient privacy concerns, there are few public medical datasets available for querying, leading to a limited and homogeneous transfer dataset.
Second, existing methods primarily target image classification, where every class prediction from the victim model is useful. In contrast, medical text generation involves a much larger output space (i.e., vocabulary), with only a subset of tokens being medically relevant and valuable for training a medical model.

While Knockoff Nets~\citep{knockoff} showed that a diabetic image classification model can be stolen using the non-medical ImageNet dataset, we demonstrate that such images are not suitable for stealing medical \acp{MLLM}. When using non-medical images, medical \acp{MLLM} produce simplistic, repetitive reports, with few containing relevant disease.
This underscores the importance of aligning the query distribution with the victim model's data distribution for successful model stealing, as highlighted in previous works~\cite{knockoff, dfme, zhuang2024stealthy}.


To address these problems, we introduce \ac{ADA-Steal}, the first data-free method for stealing medical \acp{MLLM} focused on radiology report generation. This method addresses the issues by: (i) using an open-source oracle model to diversify reports without requiring prior medical knowledge, and (ii) integrating the new reports into the query images via targeted adversarial attacks. This process enables non-medical query data to produce more diverse and medically relevant reports, effectively aligning the non-medical domain data with the medical domain data.

\paragraph{Contributions.}
(i) We are the first to investigate the feasibility of model stealing attacks against medical \acp{MLLM} and to identify the associated challenges.
(ii) We propose \acf{ADA-Steal}, the first model theft method to replicate the functionality of medical \acp{MLLM} without requiring expert knowledge or access to medical domain data.
(iii) We validate our \ac{ADA-Steal} method on the \textsc{IU X-Ray} and \textsc{MIMIC-CXR} test datasets, showing that it approaches the victim model's performance in both natural language generation metrics and clinical efficacy metrics, even when using the non-medical \textsc{CIFAR100} dataset.
(iv) We conduct ablation studies to analyze the effects of different components of our method, showing that \ac{ADA-Steal} can increase the diversity of the attacker dataset by medical report enrichment and domain alignment.

\section{Related Work}
\paragraph{Knowledge distillation.}
Knowledge distillation~\citep{model-compression, kd} helps transfer the knowledge from a complex and larger ``teacher'' model to a compact and simpler ``student'' model, which is similar to our victim-attacker design. 
However, unlike knowledge distillation, where the student model has the same data distribution as the teacher model's training data, in our problem formulation, the attacker has no prior knowledge of the victim's black-box model, e.g., unknown architecture, data distributions or training parameters. 
Although data-free knowledge distillation~\citep{dfad, zero-kd, dfkd} further assumes the absence of the teacher model's training data, its requirement of white-box access to the teacher model for backpropagation is a major difference to our setup.
\paragraph{Model stealing.}
Model stealing, or model theft, typically has one of two objectives: exact replication of the model or its components, or functionality replication, where the attacker aims to mimic the model's behavior.
The first type focuses on extracting the model's hyperparameters~\citep{wang2018stealing}, architecture~\citep{joon18iclr}, or learned parameters~\citep{steal-api}. The second type~\citep{knockoff, dfme, zhuang2024stealthy}, involves training a model that mimics the victim's performance without prior knowledge of its training data or architecture.
In this work, we present the first functionality model stealing attack against \acp{MLLM} for radiology report generation. While previous methods like Knockoff Nets~\citep{knockoff} replicate medical image classifiers using natural images, they deal with a much smaller output space compared to text generation. Bert-Thieves~\citep{krishna2019thieves}, another related approach, targets language models, but does not handle images and benefits from publicly available text data that shares a similar distribution with the victim model.
In contrast, medical data is hard to obtain, and only a specific subset of the vocabulary is relevant in this context, making it more challenging for the attacker, who may lack prior knowledge in the medical domain.
\paragraph{Security of \aclp{MLLM}.}
With the ability to understand and reason about different data types, \acp{MLLM} are vulnerable to evasion attacks targeting each data modality, such as malicious image and text constructs~\citep{mllm-aa, mllm-safety}.
In contrast to these attacks that are designed to induce erroneous or disallowed responses, the model stealing attack we present aims to mimic the functionality of the victim medical \ac{MLLM} for radiology report generation, using only black-box access to the victim model.

\begin{figure*}[t]
\centering
\includegraphics[width=0.9\textwidth]{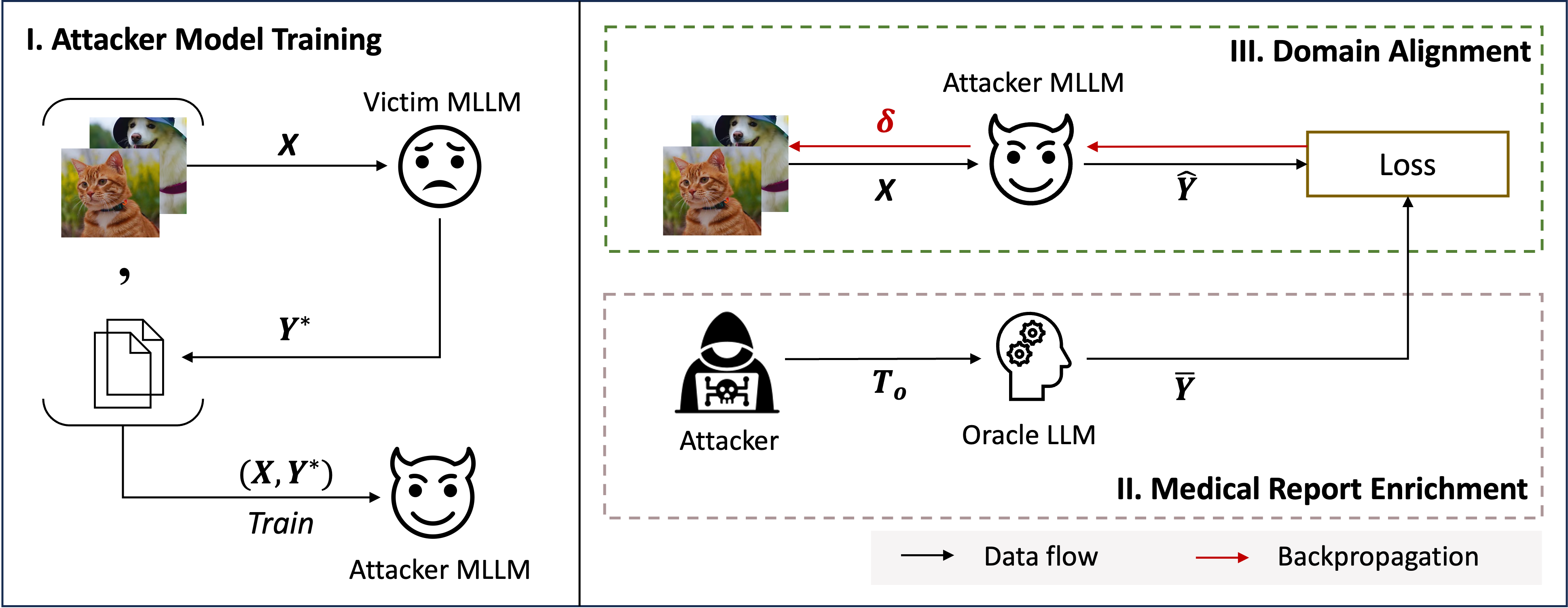} 
\caption{The overview of our proposed approach with three iterative phases: (I) attacker model training, (II) medical report enrichment (in gray dash box), and (III) domain alignment (in green dash box).}
\label{fig: method}
\end{figure*}

\section{Threat Model}
In this section, we formalize the threat model for stealing black-box medical \acp{MLLM} in radiology report generation.
First, we introduce preliminary concepts and notations.
We then formalize the victim model as well as the attacker's objective and knowledge based on the real-world setup.

\paragraph{Notations.}
We model \acp{MLLM} predicting the probability of the next token $y_l$ given the preceding language tokens $y_{<l}$, the input image $X$, and the instruction prompt $T$. The final output is the answer $Y=\{y_l\}_{l=1}^L$. Each token $y$ is drawn from vocabulary $\mathbb{V}$. This is formalized as:

\begin{equation}\label{eq:language_modelling}
    p(Y|X, T)=\prod_{l=1}^L p(y_l|y_{<l}, X, T),
\end{equation}

A table summarizing all notations used in this paper is provided in Appendix A.

\paragraph{Victim model.} 
The victim medical model $M_v$ with parameters $\theta_v$ is an \ac{MLLM} developed for automated medical image interpretation. It accepts the image $X$ and instruction prompt $T$ as inputs, and outputs the answer $Y^* = M_v(X, T; \theta_v)$. We refer to $Y^*$ as the pseudo-report. We consider a deterministic victim making predictions based on beam search.
The model can perform various clinical tasks depending on the prompt $T$, such as image view classification, disease classification, and radiology report generation. We focus primarily on radiology report generation. This task has a much larger output space ($|\mathbb{V}_v|$) than classification tasks, and valid medical reports typically only use a small portion of the entire vocabulary. This characteristic poses additional challenges for model stealing. For simplicity, we fix the prompt $T$ for the radiology report generation task and vary only the input image $X$. We use $Y^* = M_v(X)$ as a shorthand for victim predictions under fixed $T$. The original victim training dataset is denoted as $\mathcal{D}_v=\{(X_v,Y_v)\}$, where $X_v \sim P_v$ (i.e., data distribution of the victim training images) and $Y_v$ is its corresponding medical report. 

\paragraph{Goal and knowledge of the attacker.}
The attacker aims to train a surrogate model $M_a(X, T; \theta_a)$, parameterized by $\theta_a$, that is able to generate the radiology report $\hat{Y}$ from the image $X$ similar to those produced by the victim model $M_v$ with instruction prompt $T$.
Here as well we consider $T$ to be fixed for this specific task. The attacker is allowed to query the victim model with any image and receive the corresponding report. However, the attacker lacks knowledge of: (i) the internals of the victim medical \ac{MLLM} $M_v$, including its architecture; (ii) the dataset $\mathcal{D}_v$ used to train the victim model; (iii) the vocabulary $\mathbb{V}_v$ of $M_v$; and (iv) the probability distribution of each token in the victim output $Y^*$.

\section{Adversarial Domain Alignment}
The attacker aims to replicate the radiology report generation functionality of a black-box medical model without access to medical datasets. To achieve this, they would ideally optimize the parameters $\theta_a$ of the attacker model $M_a$ to minimize the token prediction loss $\mathcal{L}$ on the victim dataset $\mathcal{D}_v$:
\begin{equation}
\label{eq:objective}
\min_{\theta_a}  \frac{1}{|\mathcal{D}_v|} \sum_{(X_v,Y_v) \in \mathcal{D}_v}\left[ \mathcal{L}(M_a(X_v),Y_v) \right]
\end{equation}

Since the attacker does not have access to $\mathcal{D}_v$, they need to construct an own dataset $\mathcal{D}_a$ for training:
\begin{equation}\label{eq:transfer-set}
    \mathcal{D}_a = \left\{(X_a, Y^*) \mid X_a \sim P_a, Y^* = M_v(X_a) \right\},
\end{equation}
where $X_a$ can be sourced from a publicly available non-medical dataset with distribution $P_a$, and $Y^*$ is the pseudo-report predicted by the victim model. 
However, using non-medical images as queries yields homogeneous reports that barely explore the output vocabulary space of the victim model.
Moreover, the attacker lacks the prior knowledge to guide the exploration and diversity of radiology reports.
To address these challenges, we propose \acf{ADA-Steal}, which initially diversifies the reports and then aligns $P_a$ with $P_v$ through data augmentation based on adversarial attacks.
In turn, this enables the attacker to obtain more varied, medically relevant reports from the victim. To this end, the objective of the attacker is:
\begin{align}
\label{eq:joint_objective}
\underset{\theta_a, \delta}{\text{minimize}} \; & \frac{1}{|\mathcal{D}_a|} \sum_{(X_a,Y^*) \in \mathcal{D}_a}\left[ \mathcal{L}(M_a(X_a),Y^*) \right. \nonumber \\
& \left. + \mathcal{L}(M_v(X_a + \delta), \Bar{Y}) \right],
\end{align}
where $\delta$ is the adversarial perturbation on image $X_a$ to elicit a more diverse and medically relevant report $\Bar{Y}$ from the victim model, which is later replaced by its proxy, the attacker model. The goal is for the perturbed query data to better approximate the distribution $P_v$.

The overall method consists of three steps: (I) \textbf{attacker model training} to mimic the victim model; (II) \textbf{medical report enrichment} to diversify the victim's pseudo-reports; and (III) \textbf{domain alignment} to shift the attacker query image distribution towards the medical image distribution. Three steps can be iterated until the query budget $B$ is exhausted. The overview of the pipeline is shown in \Cref{fig: method}.

\subsection{Attacker Model Training}
\label{subsec:attacker_model_training}
Following standard model stealing, the attacker queries the victim model with initial non-medical images from the distribution $P_a$ and receives radiology report outputs. The attacker model $M_a$ is then trained to minimize the loss in \Cref{eq:stealing_target} on the attacker's dataset $\mathcal{D}_a$:

\begin{equation}\label{eq:stealing_target}
    \mathcal{L}(M_a(X_a),Y^*) = -\sum_{l=1}^L \log p(y^*_l \mid M_a(X_a)_{<l})
\end{equation}
Once trained, this model serves as the proxy for the victim model in step III to design adversarial perturbations to be transferred to the victim. The dataset $\mathcal{D}_a$ will be iteratively updated with aligned data following step III.

\subsection{Medical Report Enrichment} \label{subsec:medical_report_enrichment}
Repetitive medical pseudo-reports from natural images limit the attacker model's ability to generalize to real medical images, which often feature varied abnormalities. Since the attacker lacks the expertise to design accurate reports, we incorporate an additional open-access large language model (LLM) as our oracle model $M_o$ to generate a more diverse and medical-relevant report $\Bar{Y}$.
Here, $\Bar{Y} \sim M_o(T_o)$, where $T_o$ is the prompt for the oracle model as follows.
The report $\Bar{Y}$ will be used in the next step as the desired output for the query image.
We emphasize that $M_o$ does not benefit from the image modality.

\begin{tcolorbox}[
    colback=white!, 
    colframe=black!, 
    colbacktitle=black!, 
    title=Prompt $T_o$,
    fontupper=\scriptsize, 
]
\small 
  Give me some examples of normal/abnormal descriptions of the airway, breathing, cardiac, diaphragm, and everything else (e.g., mediastinal contours, bones, soft tissues, tubes, valves, and pacemakers) for chest X-rays.
\end{tcolorbox}

\subsection{Domain Alignment}
\label{subsec:adversarial_domain_alignment}
Training the attacker model  $M_a $ on $ P_a $ using the victim model's predictions makes $ M_a $ a proxy for the victim model. This suggests that $ M_a $ can map $ X_v $ to $ Y_v $, even with initial low accuracy. We hypothesize that an input $ X $ can be generated to produce a relevant report $ \Bar{Y} $ by reversing the mapping, $ X = M_a^{-1}(\Bar{Y})$.
We adapt the Fast Gradient Sign Method (FGSM)~\cite{fgsm} to generate adversarial perturbations $\delta$ on image $X_a$ with the attacker model $M_a$ and the new report $\Bar{Y}$:

\begin{equation}
\label{eq:adversarial_perturbation}
    \delta = \epsilon \cdot \text{sign}(\nabla_{X_a} \mathcal{L}(M_a(X_a),\Bar{Y})),
\end{equation}
where $\epsilon$ is the magnitude of the adversarial perturbations.
The attacker can query the victim model with the generated image $X_a + \delta$ and update the attacker transfer dataset $\mathcal{D}_a$:

\begin{equation}
    \mathcal{D}_a=\{(X_a + \delta,M_v(X_a + \delta))\}.
\end{equation}

Repeat steps I-III until query budget $B$ is exhausted.

\section{Experimental Setup}
We now introduce the setup used in our experiments, including models, datasets, attacker image distribution, baseline attacks, evaluation metrics, and implementation details.

\begin{table}[t]
\centering
\begin{tabular}{@{}ccc@{}}
\toprule
\textbf{\textsc{Model}} & \textbf{\textsc{No. parameters}} & \textbf{\textsc{Role}} \\ \midrule
\textsc{CheXagent} & 8 bn & Victim \& Attacker \\
\textsc{IDEFICS} & 9 bn & Attacker \\
\textsc{Zephyr} & 7 bn & Oracle \\
\bottomrule
\end{tabular}
\caption{Models used in our experiments.}
\label{tab: models}
\end{table}

\paragraph{Models.}
To verify our proposed method, we use three pre-trained models.
\textsc{CheXagent}~\cite{chexagent} is a 7 billion-parameter medical foundation model, designed to analyze and summarize CXRs. This \ac{MLLM} is used as the victim in all our experiments.
We denote \textsc{CheXagent*} the version of the same model used by the attacker, with the vanilla \ac{LLM} weights (i.e., \textsc{Mistral}-7B) instead of the clinically fine-tuned version.
\textsc{IDEFICS}~\cite{idefics} is a 9 billion-parameter \ac{MLLM} trained on image-text pairs on various multimodal benchmarks. We directly use this pre-trained \ac{MLLM} as attack baselines.
\textsc{Zephyr}-7B~\cite{zephyr} is used as oracle \ac{LLM} for our \ac{ADA-Steal} in all experiments.
An overview of model architectures and their role in our work is provided in \Cref{tab: models}.

\begin{table}[t]
\centering
\resizebox{\columnwidth}{!}{%
\begin{tabular}{@{}cccc@{}}
\toprule
\textbf{\textsc{Dataset}} & \textbf{\textsc{Train/Test}} & \textbf{\textsc{Image}} & \textbf{\textsc{Label}} \\ \midrule
\textsc{CIFAR-100} & 50k/10k & Non-medical  & Image classes \\
\textsc{MIMIC-CXR} & 369k/5k & Chest X-ray & Radiology reports \\
\textsc{IU X-Ray} & 5k/0.8k & Chest X-ray & Radiology reports \\ \bottomrule
\end{tabular}%
}
\caption{Overview of datasets.}
\label{tab: datasets}
\end{table}

\paragraph{Datasets.}
We validate \ac{ADA-Steal} on three standard datasets: \textsc{IU X-Ray}~\cite{iux-ray}, \textsc{MIMIC-CXR}~\cite{mimic-cxr}, and \textsc{CIFAR-100}~\cite{cifar}.
\textsc{IU X-Ray} and \textsc{MIMIC-CXR} are medical datasets consisting of chest X-rays and their corresponding radiology reports.
Following previous work~\citep{r2gen, r2cmn, radialog}, we only consider the Findings section in the radiology reports and exclude samples without it.
Furthermore, MIMIC-CXR is one of the training datasets of the victim \textsc{CheXagent}~\cite{chexagent}.
\textsc{CIFAR-100} is extensively used in the computer vision community, but unrelated to the medical field.
Our goal is to show that natural images can be used to steal medical IP, despite originating from a different domain or data distribution.
\Cref{tab: datasets} summarizes the three datasets with official or conventional training and test split sizes.
\begin{table*}[t]
\centering
\begin{small}
\begin{tabular}{@{}cl| rrr | rrr@{}}
\toprule
\multicolumn{2}{c|}{\textsc{\textbf{Test Data}}} & \multicolumn{3}{c|}{\textsc{\textbf{MIMIC-CXR}}} & \multicolumn{3}{c}{\textsc{\textbf{IU X-Ray}}} \\ \midrule
\multicolumn{2}{c|}{\textsc{\textbf{Metrics}}} & \multicolumn{1}{c}{\textsc{RG-L}} & \multicolumn{1}{c}{\textsc{Bert-S}} & \multicolumn{1}{c|}{\textsc{Rad-S}} & \multicolumn{1}{c}{\textsc{RG-L}} & \multicolumn{1}{c}{\textsc{Bert-S}} & \multicolumn{1}{c}{\textsc{Rad-S}} \\ \midrule
\multicolumn{1}{c|}{\textsc{\textbf{Victim}}} & \textsc{CheXagent} & 26.5\, (1.00$\times$) & 53.0\, (1.00$\times$) & 20.7\, (1.00$\times$) & 32.2\, (1.00$\times$) & 58.7\, (1.00$\times$) & 26.1\, (1.00$\times$) \\ \midrule
\multicolumn{1}{c|}{\multirow{8}{*}{\textsc{\textbf{Attacker}}}} & \textsc{IDEFICS} & 14.6\, (0.55$\times$) & 8.0\, (0.15$\times$) & 0.7\, (0.04$\times$) & 14.1\, (0.44$\times$) & 11.2\, (0.19$\times$) & 0.3\, (0.01$\times$) \\
\multicolumn{1}{c|}{} & \textsc{ +Knockoff} & 20.2\, (0.76$\times$) & 45.3\, (0.85$\times$) & 12.5\, (0.60$\times$) & 22.9\, (0.71$\times$) & 48.1\, (0.82$\times$) & 14.0\, (0.54$\times$) \\
\multicolumn{1}{c|}{} & \textsc{ +ADA-Steal} & \textbf{23.2}\, (0.88$\times$) & \textbf{49.0}\, (0.92$\times$) & \textbf{15.6}\, (0.75$\times$) & \textbf{29.8}\, (0.93$\times$) & \textbf{52.9}\, (0.90$\times$) & \textbf{19.4}\, (0.74$\times$) \\ \cmidrule(l){2-8} 
\multicolumn{1}{c|}{} & \textsc{CheXagent*} & 10.5\, (0.40$\times$) & 0.0\, (0.00$\times$) & 0.6\, (0.03$\times$) & 6.6\, (0.20$\times$) & 0.0\, (0.00$\times$) & 0.5\, (0.02$\times$) \\
\multicolumn{1}{c|}{} & \textsc{ +Knockoff} & 23.2\, (0.88$\times$) & 43.7\, (0.82$\times$) & 16.3\, (0.79$\times$) & 26.1\, (0.81$\times$) & 48.3\, (0.82$\times$) & 22.0\, (0.84$\times$) \\
\multicolumn{1}{c|}{} & \textsc{ +ADA-Steal} & \textbf{24.7}\, (0.93$\times$) & \textbf{44.5}\, (0.84$\times$) & \textbf{18.7}\, (0.90$\times$) & \textbf{26.6}\, (0.83$\times$) & \textbf{52.5}\, (0.89$\times$) & \textbf{25.8}\, (0.99$\times$) \\ \bottomrule
\end{tabular}
\end{small}
\caption{Performance metrics on test data (best value in \textbf{bold}, followed by the ratio to the original victim performance).} 

\label{tab:main-results}
\end{table*}
\paragraph{The attacker query data $P_a$.}
The attacker queries the victim with images from a large discrete image distribution $P_a$, as shown in \Cref{eq:transfer-set}.
The experiments differ in four choices of $P_a$, each of which is explained below.
\begin{itemize}
\item $P_a=$ \textsc{CIFAR-100}: we sample natural images from the training set of \textsc{CIFAR-100} as the attacker query data.

\item $P_a=$ MIMIC-CXR: we assume the attacker has access to the radiographs in the training set of the MIMIC-CXR as the initial query data.
This setup serves as an upper bound on the stealing performance the attacker can achieve.

\item $P_a\sim\mathcal{N}(0,1)$: images are randomly generated with pixel values following a standard normal distribution, scaled to [0, 255], and rounded to the nearest integer.

\item $P_a=$ mix of \textsc{CIFAR-100} and MIMIC-CXR: we include a variable number of images from the victim training set (i.e., images from MIMIC-CXR) into the attacker dataset. The proportion of these images is controlled by the ratio $r$, defined as the percentage of \textsc{MIMIC-CXR} images within the initial attacker dataset.
\end{itemize}

\paragraph{Attacks.}
We first test the original performance of \textsc{IDEFICS} and \textsc{CheXagent*} on the medical dataset as a baseline reference, followed by evaluating two attack strategies.
\textsc{Knockoff} fine-tunes the attacker model using the method of Knockoff Nets~\cite{knockoff} with the same \ac{MLLM}.
\ac{ADA-Steal} denotes our proposed method.

\paragraph{Evaluation metrics.}
We evaluate our attacker model performance on the test sets of \textsc{IU X-Ray} and \textsc{MIMIC-CXR}.
After preprocessing, the number of test samples in \textsc{MIMIC-CXR} and \textsc{IU X-ray} is 3858 and 590, respectively.
Following \textit{CheXbench}, a benchmark designed by~\citet{chexagent} to evaluate models across eight CXR interpretation tasks (e.g., radiology report generation).
We evaluate the above models by two types of metrics:
the natural language generation metrics include ROUGE-L (\textsc{RG-L})~\cite{rouge} and BERT-Score (\textsc{Bert-S})~\cite{bert-score}, while the clinical efficacy metrics include RadGraph-Score (\textsc{Rad-S})~\cite{radgraph} and GPT-4 evaluation.
In particular, the \textsc{Rad-S} assesses the quality of generated radiology reports by employing output reports to construct  RadGraphs to identify entities and their relations in comparison to ground truth references.
The performance on the CheXbert metric is not reported due to its unreliability for out-of-distribution reports.
Detailed explanations and the prompt for GPT-4 evaluation are included in Appendix C and D, respectively.

\paragraph{Implementation details.}
We initialize the attacker query set with 500 images from CIFAR-100, and then repeat the steps of our method three times, resulting in a total query budget of $B=1500$.
We set the probabilities of abnormal, normal, and original ($\hat{Y}$) anatomical descriptions into 80\%, 10\%, and 10\%, respectively, in the new report $\Bar{Y}$ for adversarial perturbation generation.
The learning rates for fine-tuning \textsc{IDEFICS} and \textsc{CheXagent*} are fixed to $5\times10^{-6}$ and $1\times10^{-5}$, respectively, without weight decay.
The maximum new sequence length is set to 512, and a diversity penalty of 0.2 with three beam groups, each containing six beams, is applied. 
The top-1 response is collected as the generated report.
The adversarial noise budget $\epsilon$ is set to $0.2$ unless otherwise specified. 
In both the victim and attacker models, the report generation process employs the same prompt $T$, as shown below.
All experiments are conducted on a single NVIDIA A100 GPU.

\begin{tcolorbox}[
    colback=white!, 
    colframe=black!, 
    colbacktitle=black!, 
    title=Fixed prompt $T$ for victim and attacker model,
    fontupper=\scriptsize, 
]
\small 
  Write a structured Findings section for the given image as if you are a radiologist.
\end{tcolorbox}

\section{Experimental Results}
\paragraph{Model stealing performance.}
The main experimental results on the two aforementioned test datasets are shown in \Cref{tab:main-results}.
First, both attack strategies improve the performance of radiology report generation compared to the original \ac{MLLM}.
This outcome highlights the vulnerability of the medical \acp{MLLM} to model stealing attacks.
Second, \ac{ADA-Steal} outperforms the other attack in all metrics, which confirms the effectiveness of our proposed model stealing method in enhancing the diversity of the attacker set.
Such diversity may come from the incorporation of medical report enrichment and adversarial image generation that align attacker predictions with the expert knowledge of the oracle LLM.
Third, when comparing between datasets, the performance achievements of \ac{ADA-Steal} tested on \textsc{IU X-Ray} are higher than that of \textsc{MIMIC-CXR}.
This might be because \textsc{IU X-Ray} is relatively small and has less diverse image-text mappings, making it easier for attackers to mimic the image-to-text generation functionality.
Finally, there is no clear winner in terms of the model architecture used by the attacker, but rather the performance depends on both the attack strategy and the measured metric.

\begin{figure*}[ht]
  \centering
  \begin{minipage}[b]{0.48\textwidth}
        \centering
        \includegraphics[width=\textwidth]{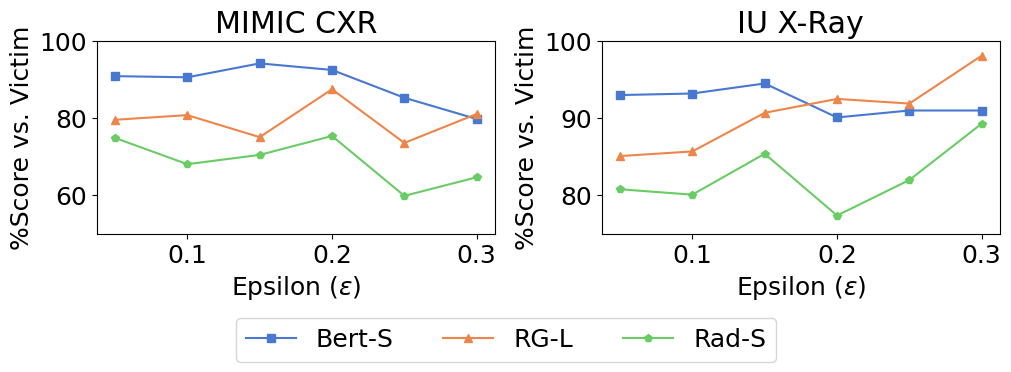}
        \subcaption{Stealing performance of the adversarial noise budget $\epsilon$.} 
        \label{fig:evaluations}
    \end{minipage}
    \hfill
    \begin{minipage}[b]{0.48\textwidth}
        \centering
        \includegraphics[width=\textwidth]{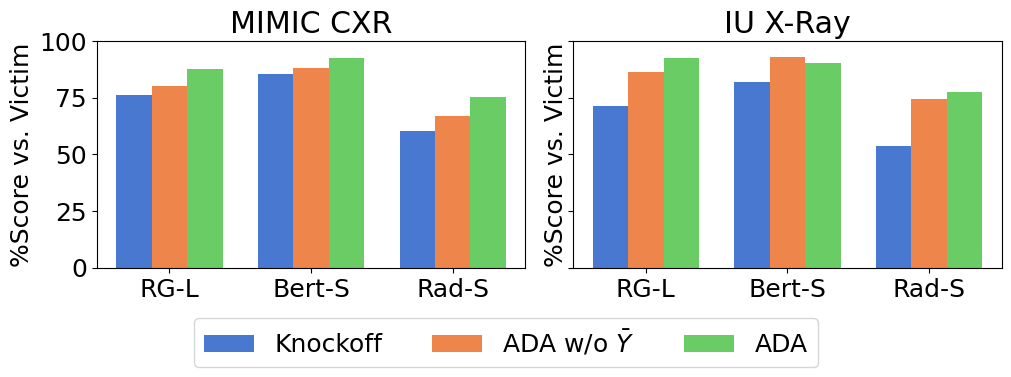}
        \subcaption{Ablation performance of \ac{ADA-Steal} without Oracle.} 
        \label{fig:ablation}
    \end{minipage}
  \caption{Left: performance of \ac{ADA-Steal} on \textsc{IDEFICS} with different $\epsilon$ across three metrics compared to the victim. Right: ablation study compares the performance of three attackers.}
  \label{fig: epsilon}
\end{figure*}
\paragraph{Ablative analysis.}
To investigate the effect of adversarial attacks in domain alignment as well as oracle LLM diversification in medical report enrichment, an abrogation study is conducted, examining the impact of the non-utilization of the oracle model in \ac{ADA-Steal} (denoted as ADA w/o $\Bar{Y}$). 
Under the same experimental setup of \textsc{IDEFICS} and $\epsilon=0.2$, we compare its model performance to \textsc{Knockoff} and \ac{ADA-Steal} and show the results in  \Cref{fig:ablation}.
First, ADA w/o $\Bar{Y}$  always outperforms \textsc{Knockoff}, confirming the added value of adversarial images toward domain alignment.
While the victim model tends to produce simple, repetitive pseudo-reports with no disease indications, the adversarial images generated during domain alignment enhance the confidence and quality of their pseudo-reports.
Second, applying the oracle model to diversify the pseudo-reports can further improve the model stealing ability.
This observation is consistent with our hypothesis that the medical report enrichment helps increase the diversity and maintain the clinical relevance of adversarial images by aligning the attacker's predictions with its own knowledge.

\begin{figure}[t]
\centering
\includegraphics[width=\columnwidth]{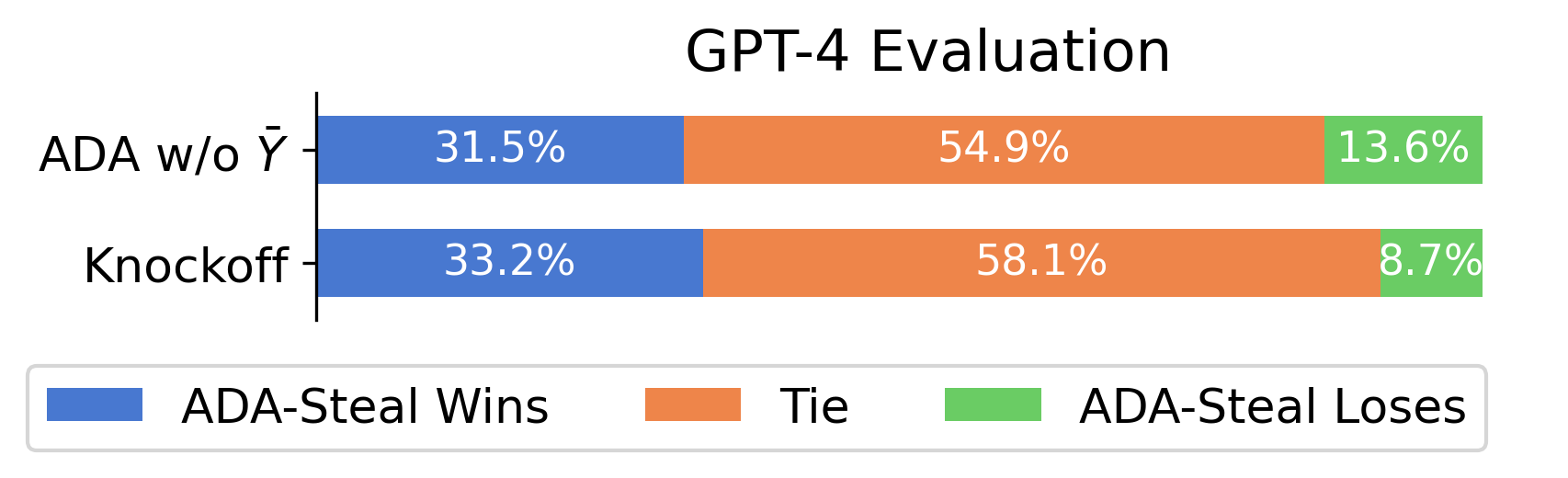} 
\caption{Qualitative evaluation by GPT-4 comparing the quality of test reports pairwise.} 
\label{fig:gpt4}
\end{figure}


\paragraph{Adversarial noise budget $\mathbf{\epsilon}$.}
To analyze the impact of the amount of adversarial noise introduced, we train our attacker model (\textsc{IDEFICS+ADA-Steal)} for different $\epsilon$ in the range from 0.05 to 0.3.
\Cref{fig: epsilon} shows the results on \textsc{MIMIC-CXR} and \textsc{IU X-Ray}.
For all performance metrics, the general trend of the score is to rise and then fall as $\epsilon$ increases.
This pattern can be attributed to the intrinsic nature of the perturbation. 
An extremely small $\epsilon$ introduces insufficient perturbation to alter the original image significantly. 
Conversely, the perturbation resulting from a comparatively large $\epsilon$ can push pixel values out of the data distribution.
While there is not one optimal value for $\epsilon$ across datasets and metrics, we note that the range between 0.05 to 0.2 yields overall good performance.

\begin{table}[t]
\centering
\resizebox{\columnwidth}{!}{%
\begin{tabular}{@{}c|rrr|rrr@{}}
\toprule
\multirow{2}{*}{$P_a(X)$} & \multicolumn{3}{c|}{\textbf{\textsc{MIMIC-CXR}}} & \multicolumn{3}{c}{\textbf{\textsc{IU X-Ray}}} \\
 & \textsc{RG-L} & \textsc{Bert-S} & \textsc{Rad-S} & \textsc{RG-L} & \textsc{Bert-S} & \textsc{Rad-S} \\ \midrule
$\emptyset$ & 14.6 & 8.0 & 0.7 & 14.1 & 11.2 & 0.3 \\ \midrule
\textsc{CIFAR-100} & 21.4 & \textbf{48.0} & 14.1 & \textbf{27.6} & \textbf{54.6} & \textbf{22.3} \\ 
\textsc{$r=0.1$} & \textbf{24.1} & 47.9 & \textbf{15.5} & 27.4 & 53.6 & 21.1 \\
\textsc{$r=0.5$} & 20.4 & 42.8 & 10.2 & 21.0 & 42.2 & 9.5 \\ 
\textsc{MIMIC-CXR} & 18.7 & 38.3 & 7.9 & 13.8 & 30.3 & 10.4 \\ \midrule
\textsc{Random} & 7.0 & 0.0 & 0.0 & 5.3 & 0.0 & 0.0 \\ \bottomrule
\end{tabular}%
}
\caption{Evaluation of \textsc{IDEFICS+ADA-Steal} using different attacker image distributions $P_a$, where the attacker models are fine-tuned with $\epsilon=0.05$.}
\label{tab: pa}
\end{table}

\begin{figure*}
\centering
\includegraphics[width=0.9\textwidth]{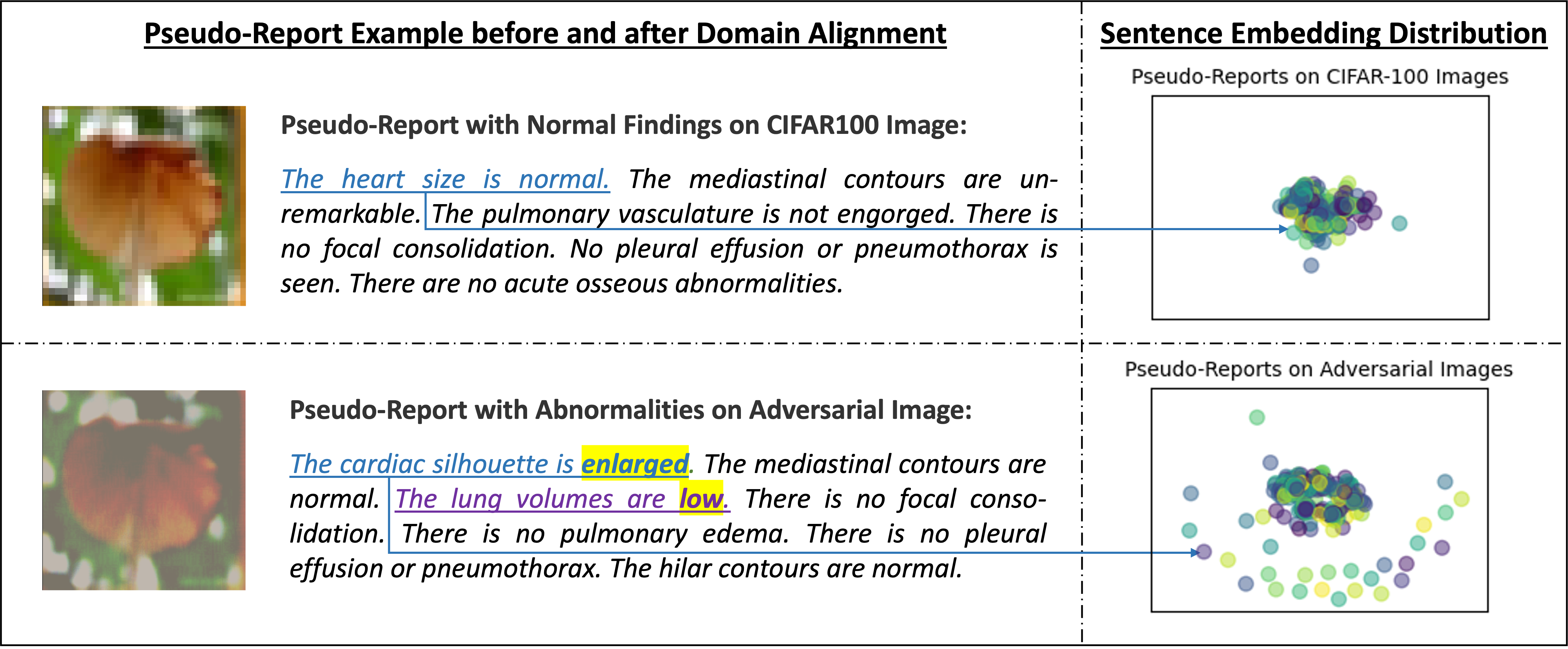} 
\caption{Left: Example of an image and pseudo-report pair from the Mushroom class in \textsc{CIFAR-100}. The abnormality is marked in yellow, with the victim model predicting an enlarged heart and low lung volumes in the adversarial pseudo-report.
Right: t-SNE visualizations of pseudo-report sentence embeddings on the \textsc{CIFAR-100} image and its adversarial counterpart from \textsc{\ac{ADA-Steal}}; showing that abnormalities increase report diversity. Arrows indicate the sentence embeddings for these descriptions.}
\label{fig:qualitative}
\end{figure*}

\paragraph{Image distribution $P_a$.}
\Cref{tab: pa} illustrates the evaluations of \textsc{IDEFICS+ADA-Steal} fine-tuned on different attacker dataset distributions.
The row $P_a=\emptyset$ shows the radiology report generation ability of the open-access \textsc{IDEFICS}-9b model.
The following observations can be made from the results.
First, initializing attacker images from the normal distribution $\mathcal{N}(0, 1)$ (row \textsc{Random}) fails to steal the model, resulting in a performance inferior to that of the original \textsc{IDEFICS}.
This failure can be explained by the significant gap between the normal distribution and the actual distribution of medical images, which causes the attacker to be inadequate in providing the necessary medical information; in this setup, our approach to enhance data diversity does not help, as random noise is diverse, but is rather lacking domain relevance.
Second, the attacker sets constructed with image sampling from the other four image distributions help improve the model's ability to generate radiology reports.
The counterintuitive observation that $P_a=$ MIMIC-CXR leads to worse performance than the other three due to the fact that the synthesized adversarial images on the actual CXRs distort the image distribution from the actual CXRs.

\paragraph{Impact of oracle and victim models.}
We further evaluate the flexibility and generalizability of our method with different oracle and victim models and record the detailed experimental performance in Appendix B. 
As shown in Figure 6, using GPT-4 as the oracle model, due to its greater capabilities, improves the performance of \ac{ADA-Steal}. 
Additionally, using \textsc{Med-Flamingo}~\cite{med-flamingo} as the victim model, Table 6 shows that \ac{ADA-Steal} outperforms baselines and even surpasses the victim model in RG-L and \textsc{Bert-S metrics}.

\paragraph{Qualitative analysis.}
We further investigate the quality of pseudo-reports generated by \textsc{\ac{ADA-Steal}} based on augmented images.
\Cref{fig:qualitative} shows an example of pseudo-reports generated for a CIFAR-100 image and for its adversarial counterpart used in \textsc{\ac{ADA-Steal}} (left).
The images in the attacker queries constructed with \textsc{\ac{ADA-Steal}} tend to induce more diverse pseudo-reports which include descriptions of abnormalities, while the other attacks fail to do so.
This phenomenon is consistent throughout the generated data and aligns with our design objective of the oracle module meant to induce more anomalies.
The t-SNE~\citep{t-sne} visualization of the pseudo-report sentence embedding space (\Cref{fig:qualitative}, right) further confirms the diversity introduced by the use of adversarial images in \textsc{\ac{ADA-Steal}}.
Finally, we use \mbox{GPT-4} to assess the quality of the pseudo-reports generated by \textsc{\ac{ADA-Steal}} in comparison to those from \textsc{Knockoff} and ADA w/o $\Bar{Y}$ applied to \textsc{IDEFICS}.
\Cref{fig:gpt4} shows that \ac{ADA-Steal} generates more high-quality reports than the other attacks.

\section{Discussion}
\paragraph{Extended scope.}
Our current setup applies to victim models that are deterministic and use beam search decoding.
We see no fundamental limitation in our methodology that would prevent us from extending our work to models with stochastic outputs or that vary the output decoding strategy.

\paragraph{Reports' drift.}
The reports generated by the oracle LLM are only based on a prompt and no images. We identify a risk that these reports are not accurate or of good quality.
However, we notice in practice that data diversity close to the target domain is sufficient to elicit data diversity in the victim model.
Moreover, the reports are not used to train the stolen model, further limiting their potential impact.

\paragraph{Defenses.}
Our work does not evaluate the \ac{ADA-Steal} against model stealing defenses. Current defenses~\citep{Orekondy2019PredictionPT, kariyappa2020defending, mazeika2022steer} typically add noise to the victim model’s predictions (i.e., logits) to hinder the attacker while maintaining the model's utility (e.g., top-class predictions). However, these defenses are not suited to the complex input and output spaces addressed in this paper. Additionally, in the medical field, it is crucial that reports generated by medical \acp{MLLM} remain truthful and accurate.
As such, defending against the present attack is not trivial and would require dedicated defenses.

\paragraph{Ethical considerations.}
This paper advances research on model stealing attacks, emphasizing the significance of this threat and the urgent need for robust defenses. By raising awareness of the latest security challenges, we aim to help the community better prevent or mitigate such attacks. Our work is built entirely on open resources, and we hope it motivates further study of practical attacks on machine learning to ultimately develop safer, more reliable systems.

\section{Conclusion}
In this paper, we show for the first time that an attacker can successfully steal the functionality of radiology report generation from a medical \ac{MLLM} without access to the victim data distribution.
Our attack \ac{ADA-Steal} produces a diverse dataset for stealing by leveraging adversarial attacks for domain alignment and an oracle model for report enrichment.
Experimental results on two medical datasets demonstrate the effectiveness of \ac{ADA-Steal}, which outperforms existing methods directly adapted to the task of radiology report generation.
We encourage medical \ac{MLLM} owners to consider the risk of model theft and to protect their assets.

\section*{Acknowledgements}
We acknowledge the support and funding by Bosch AIShield. 
This work was supported by ELSA – European Lighthouse on Secure and Safe AI funded by the European Union under grant agreement No. 101070617.
This work was also granted access to the HPC resources of IDRIS under the allocations AD011013704R1, AD011011631R2, and AD011011631R4 made by GENCI.

\bibliography{aaai25}
\clearpage
\appendix

\section{A. Notations}
\Cref{tab:notation} summarizes the notations used in our paper.
\begin{table}[ht]
\centering
\resizebox{\columnwidth}{!}{%
\begin{tabular}{@{}cl@{}}
\toprule
\multicolumn{1}{c}{\textbf{\textsc{Symbol}}} & \multicolumn{1}{l}{\textbf{\textsc{Explanation}}} \\ \midrule
$X$ & A general input image \\
$X_a$ & An attacker image \\
$T$ & Fixed prompt to query victim and attacker model \\
$T_o$ & Prompt instruction for the oracle model \\
$Y^*$ & Pseudo-report given by the victim MLLM \\
$\hat{Y}$ & Reports generated by the attacker MLLM \\
$\Bar{Y}$ & Diversified reports \\ 
$M_v$ & The victim MLLM \\
$M_a$ & The attacker MLLM \\
$M_o$ & The oracle LLM \\
$\theta_v$ & Victim model parameters \\
$\theta_a$ & Attacker model parameters \\
$\mathbb{V}_v$ & Vocabulary of the victim MLLM \\
$\mathcal{D}_v$ & Victim training set \\
$\mathcal{D}_a$ & Attacker training set \\
$P_a$ & Attacker image distribution \\
$P_v$ & Victim training image distribution \\
$\delta$ & Adversarial perturbation \\
$\epsilon$ & Magnitude of adversarial perturbation \\
$B$ & Attacker query budget \\
\bottomrule
\end{tabular}%
}
\caption{Notations used in this paper.}
\label{tab:notation}
\end{table}

\section{B. Impact of Oracle and Victim Models}
This section presents two additional experiments to demonstrate the flexibility and generalizability of our proposed \ac{ADA-Steal}.

\begin{figure}[ht]
    \centering
    \includegraphics[width=0.48\columnwidth]{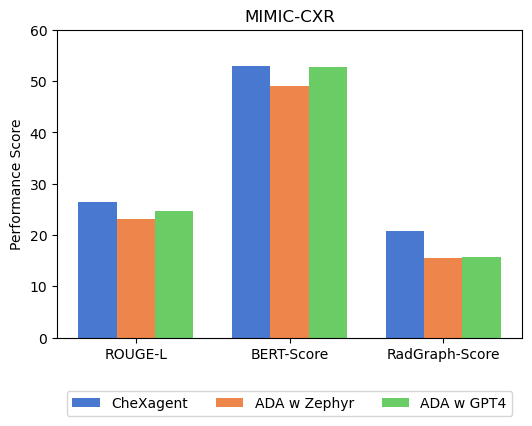}
    \includegraphics[width=0.48\columnwidth]{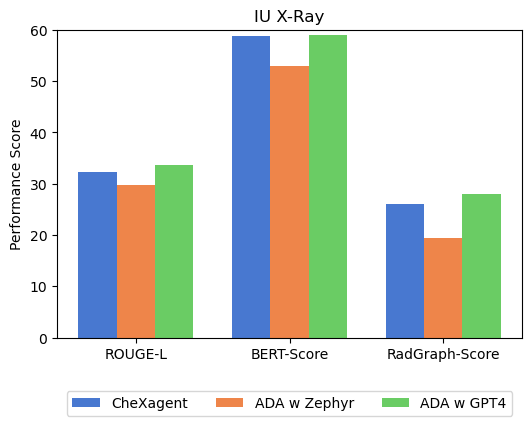}
  \caption{Performance of \ac{ADA-Steal} with \textsc{Zephyr} and GPT-4 as the oracle model.}
  \label{fig: oracle-gpt}
\end{figure}

\Cref{fig: oracle-gpt} shows the ability of our method with a more powerful oracle model, GPT-4. 
Compared to using \textsc{Zephyr-7B}, the improved performance on both the MIMIC-CXR and \textsc{IU X-Ray} datasets emphasize the effectiveness of our design in enriching the medical report with an oracle LLM.
We also evaluate our method using \textsc{Med-Flamingo}~\cite{med-flamingo} as the victim model. 
As shown in \Cref{tab:victim-medflamingo}, \ac{ADA-Steal} excels the baseline \textsc{Knockoff} and even outperforms the victim model performance in terms of natural language generation metrics, including RG-L and \textsc{Bert-S}.

\begin{table}[ht]
\centering
\resizebox{\columnwidth}{!}{%
\begin{tabular}{@{}l|r|rrr@{}}
\toprule
\multicolumn{1}{c|}{\multirow{2}{*}{}} & \multicolumn{1}{c|}{\textbf{\textsc{Victim}}} & \multicolumn{3}{c}{\textbf{\textsc{Attacker}}} \\
\multicolumn{1}{c|}{} & \multicolumn{1}{c|}{\textsc{Med-Flamingo}} & \multicolumn{1}{c}{IDEFICS} & \multicolumn{1}{c}{\textsc{+Knockoff}} & \multicolumn{1}{c}{\textsc{+ADA}} \\ \midrule
\textbf{\textsc{RG-L}} & 12.8\, (1.00$\times$) & 14.6\, (1.14$\times$) & 16.8\, (1.31$\times$) & \textbf{17.2}\, (1.34$\times$) \\
\textbf{\textsc{Bert-S}} & 20.9\, (1.00$\times$) & 8.0\,(0.38$\times$) & 22.2\, (1.06$\times$) & \textbf{23.4}\, (1.12$\times$) \\
\textbf{\textsc{Rad-S}} & 2.7\, (1.00$\times$) & 0.7\, (0.26$\times$) & 0.9\, (0.33$\times$) & \textbf{1.2}\, (0.44$\times$) \\ \bottomrule
\end{tabular}%
}
\caption{Performance with \textsc{Med-Flamingo} as the victim.}
\label{tab:victim-medflamingo}
\end{table}

\section{C. Prompt for GPT-4 Evaluation}
The prompt for GPT-4 evaluation, shown in the text box, is adapted from \textit{CheXbench} as provided by \citet{chexagent}.

\begin{tcolorbox}[
    colback=white!, 
    colframe=black!, 
    colbacktitle=black!, 
    title=Prompt for GPT-4 Evaluation,
    fontupper=\scriptsize, 
]
 \small
  You are a professional radiologist whose job is to write radiology reports based on given chest x-ray images. Your goal is to assist users by providing professional knowledge in radiology reports. \\
\lbrack Reference Report\rbrack \\
\lbrack reference\rbrack \\
\lbrack End of Reference Report\rbrack \\
\\
\lbrack Assistant 1\rbrack \\
\lbrack report1\rbrack \\
\lbrack End of Assistant 1\rbrack \\
\\
\lbrack Assistant 2\rbrack \\
\lbrack report2\rbrack \\
\lbrack End of Assistant 2\rbrack \\
\\
\lbrack Requirements\rbrack \\
1. The length of the reports is not important. \\
2. The style of the reports is not important. \\
3. The clinical accuracy is important, especially for positive findings (i.e., diseases). \\
Therefore, please focus on clinical accuracy instead of length and style. \\
\lbrack End of Requirements\rbrack \\
\\
Please compare the accuracy of their generated reports. You should tell me whether Assistant 1 is “better than”, “worse than”, or “equal to” Assistant 2. Please first compare the generated reports with the reference report to analyze which one is more in line with the given requirements. In the last line, please output a single line containing only a single label selecting from “Assistant 1 is better than Assistant 2”, “Assistant 1 is worse than Assistant 2”, and “Assistant 1 is equal to Assistant 2”.
\end{tcolorbox}

\begin{table*}[ht]
\centering
\begin{tabular}{@{}l|c|ccccc|r@{}}
\toprule
 & \multicolumn{6}{c|}{\textbf{\textsc{CheXbert F1 Score}}} &  \\
\multirow{-2}{*}{\textbf{\textsc{Pathology}}} & \textsc{Victim} & \textsc{Knockoff} & \textsc{ADA-Steal} & \textsc{The} & \textsc{Empty} & \textsc{imdb} & \multirow{-2}{*}{\textbf{\textsc{Support}}} \\ \midrule
Enlarged Cardiomediastinum & 14.8 & \textbf{6.5} & 0.0 & 0.0 & 0.0 & 0.3 & 1124 \\
Cardiomegaly & 65.4 & 7.0 & \textbf{18.4} & 2.2 & 0.0 & 0.1 & 1596 \\
Lung Opacity & 21.7 & 0.0 & 0.0 & 0.0 & 0.0 & 0.1 & 1417 \\
Lung Lesion & 12.4 & 0.0 & 0.0 & 0.0 & 0.0 & 0.0 & 285 \\
Edema & 55.5 & \textbf{0.9} & 0.0 & 0.0 & 0.0 & 0.4 & 893 \\
Consolidation & 14.1 & 0.0 & \textbf{0.5} & 0.0 & 0.0 & \textbf{0.5} & 367 \\
Pneumonia & 12.1 & 0.0 & 0.0 & 0.0 & 0.0 & 0.0 & 503 \\
Atelectasis & 19.8 & 0.2 & 0.2 & 0.0 & 0.0 & \textbf{1.6} & 1214 \\
Pneumothorax & 27.9 & 0.0 & 3.2 & 0.0 & 0.0 & \textbf{3.3} & 100 \\
Pleural Effusion & 69.3 & 0.0 & \textbf{20.2} & 0.0 & 0.0 & 0.0 & 1373 \\
Pleural Other & 0.0 & 0.0 & 0.0 & 0.0 & 0.0 & 0.0 & 184 \\
Fracture & 1.6 & 0.0 & 0.0 & 0.0 & 0.0 & 0.0 & 236 \\
Support Devices & 62.5 & 37.5 & 0.3 & \textbf{51.3} & 0.0 & 7.0 & 1332 \\
No Finding & 23.7 & 11.3 & \textbf{11.9} & 11.6 & 11.5 & 11.8 & 236 \\ \midrule
Micro Average & 43.2 & 10.1 & 7.6 & \textbf{17.1} & 3.2 & 3.4 & 10860 \\
Macro Average & 28.6 & 4.5 & \textbf{7.9} & 4.7 & 0.8 & 1.8 & 10860 \\
Weighted Average & 39.3 & 6.6 & 5.6 & \textbf{6.9} & 0.3 & 1.5 & 10860 \\
Sample Average & 37.4 & 10.1 & 8.5 & \textbf{17.2} & 5.3 & 4.5 & 10860 \\ \bottomrule
\end{tabular}%
\caption{\textsc{CheXbert} F1 scores of 14 clinical observations. "Support" denotes the frequency of each class in the ground truth.}
\label{tab:chexbert-category}
\end{table*}

\begin{table}[ht]
\centering
\resizebox{\columnwidth}{!}{%
\begin{tabular}{@{}c|cccc|c@{}}
\toprule
 & \multicolumn{4}{c|}{\textsc{\textbf{CheXbert-Score}}} &  \\ \cmidrule(lr){2-5}
 & \multicolumn{1}{c|}{} & \multicolumn{3}{c|}{\textsc{Macro Average}} &  \\
\multirow{-3}{*}{\textsc{\textbf{Corpora}}}& \multicolumn{1}{c|}{\multirow{-2}{*}{\textsc{Acc.}}} & \textsc{Precision} & \textsc{Recall} & \textsc{F1} & \multirow{-3}{*}{\textsc{\textbf{Rad-S}}} \\ \midrule
\textsc{Victim} & \multicolumn{1}{c|}{29.6} & 40.1 & 31.1 & 28.6 & 20.7 \\ \midrule
\textsc{Knockoff} & \multicolumn{1}{c|}{25.6} & 11.8 & 8.4 & 4.5 & 12.5 \\
\textsc{ADA-Steal} & \multicolumn{1}{c|}{23.5} & \textbf{29.1} & 5.2 & \textbf{7.9} & 15.5 \\
\textsc{The} & \multicolumn{1}{c|}{\textbf{28.0}} & 7.2 & \textbf{14.3} & 4.7 & NaN \\
\textsc{Empty} & \multicolumn{1}{c|}{\textbf{28.0}} & 0.4 & 7.1 & 0.8 & NaN \\
\textsc{iMDb} & \multicolumn{1}{c|}{27.5} & 21.3 & 5.5 & 1.8 & NaN \\ \bottomrule
\end{tabular}%
}
\caption{Performance of clinical efficacy on MIMIC-CXR. "NaN" presents an invalid performance score.}
\label{tab:clinical-efficacy}
\end{table}

\section{D. Failure of \textsc{CheXbert}}
In this section, we explain why we exclude the CheXbert-Score from our clinical efficacy metrics.
The CheXbert-Score uses the pre-trained \textsc{CheXbert} model~\cite{chexbert} as a multi-label classifier to assess radiology reports by performing a report labeling task, which identifies clinically important observations relative to the ground truth references. 
It evaluates 14 observations (detailed in \Cref{tab:chexbert-category}), assigning each to one of the four categories: blank, positive, negative, and uncertain.
For the 14th observation, ``\textit{no finding}'', the classification options are restricted to either blank or positive~\cite{irvin2019chexpert}. 

\Cref{tab:clinical-efficacy} records the CheXbert-Score and RadGraph-Score of six text corpora.
\textsc{Victim} (i.e., the victim \textsc{CheXagent}), \textsc{Knockoff}, and \textsc{ADA-Steal} corpora consists of radiology reports generated by these models on the images in MIMIC-CXR test set, respectively.
The texts in the other three sets are not relevant to the medical domain.
\textsc{The} has only the repetitive single word ``The'', padded to the same length as each of the ground truth reports.
\textsc{Empty} is constructed with empty strings.
\textsc{iMDb} contains texts taken at random from the Large Movie Review Dataset \cite{imdb}.
The number of texts in each corpus is the same.

The F1 score is usually taken as an indicator of CheXbert-Score performance, and \ac{ADA-Steal} does indeed excel by this measure.
However, unexpectedly, \textsc{The} outperforms \textsc{Knockoff}, rendering the results from this indicator inconclusive and not convincing.
Regarding the other indicators, \textsc{The} and \textsc{Empty} surprisingly achieve the highest accuracy score, while \ac{ADA-Steal} performs the worst.
This is due to the sparse output space of the multi-label classification task (14 categories of clinical observations, each with four or two possible
 labels) and the unbalanced nature of the test set.
Despite their non-medical focus, the \textsc{CheXbert} classifier still assigns labels to these texts.
\Cref{tab:chexbert-category} illustrates the F1 score for each of the 14 clinical observations, revealing that \textsc{The} even outperforms the radiology reports from \textsc{Knockoff} and \ac{ADA-Steal} in many respects, with a tendency to identify ``\textit{support devices}''.
Similarly, the F1 score of the \textsc{Empty} corpus primarily attributes to its success in labeling ``\textit{no finding}''.
Based on the above observations and analysis, we conclude that the CheXbert-Score is unreliable for evaluating the out-of-distribution texts and, therefore, decide not to use it as one of our clinical efficacy metrics.
\end{document}